\documentclass{article}
\begin{document}
\def\radhz{\mbox{rad}/\mbox{Hz}^{1/2}}

\title{Atmospheric gravity perturbations measured by ground-based
interferometer with suspended mirrors}

\author{V N Rudenko\dag, A V Serdobolski\dag\ and K Tsubono\ddag}

\begin{abstract}
A possibility of geophysical measurements using the large scale
laser interferometrical gravitational wave antenna is discussed.
An interferometer with suspended mirrors can be used as a
gradiometer measuring variations of  an angle between gravity
force vectors acting on the spatially separated suspensions. We
analyze restrictions imposed by the atmospheric noises on
feasibility of such measurements. Two models of the atmosphere
are invoked: a quiet atmosphere with a hydrostatic coupling of
pressure and density and a dynamic model of moving region of the
density anomaly (cyclone). Both models lead to similar
conclusions up to numerical factors.   Besides the  hydrostatic
approximation, we use a model of turbulent atmosphere  with the
pressure  fluctuation spectrum $\sim f^{-7/3}$ to explore the
Newtonian noise in a higher frequency domain (up to 10 Hz)
predicting the gravitational noise background for modern
gravitational wave detectors. Our estimates show that this could
pose a serious problem for realization of such projects. Finally,
angular fluctuations of spatially separated pendula are
investigated via computer simulation for some realistic
atmospheric data giving the level estimate $\sim
10^{-11}~$rad$\cdot\mbox{Hz}^{-1/2}$ at frequency $\sim 10^{-4}$
Hz. This looks promising for the possibility of the measurement of
weak gravity effects such as Earth inner core oscillations.
\end{abstract}

\section*{Introduction}

A possibility of geophysical measurements with large scale laser 
interferometrical gravitational wave antenna \cite{c1,c2} was discussed
in \cite{c3,c4}. It was supposed that at very low frequencies
$10^{-3}-10^{-5}$ Hz the
interferometer with suspended mirrors can be considered as an angular gravity
gradiometer measuring variations of the mutual angle between gravity force
vectors or plumb lines of spatially separated suspensions.
Geophysical information could be read out as error signal of the feedback
circuits which preserves the operational angle position of the mirrors.

As an example of geophysical phenomenon for the measurement a very weak
effect of the Earth inner core oscillations (one of ``hot points''
of modern geophysics \cite{c5}) was considered in \cite{c4}.

A number of principal instrumental problems were analyzed in  \cite{c6,c7}
(such as a necessity of suspension for the laser injection bench,
the problem of tilt and shift differentiation for spherical mirrors, and
others); the  main part of instrumental noises at low frequencies
was estimated in \cite{c7} where a positive conclusion was drawn in favor of a
feasibility of geophysical measurements. However, environmental geophysical 
fluctuations produce a predominant noise background which can be a 
problem while realizing this program.
It is clear that, for any gravitational device, the fundamental and 
unavoidable source of noises is the gravitational attraction of stochastically 
moving surrounding masses, the so-called ``Newtonian gravity noise.'' 
Such noise produced by acoustic waves, propagating under the interferometer 
base ground, was calculated in \cite{c8} at high frequency $f\geq$10 Hz typical
for gravitational detectors.

In this paper, we estimate, at least at the order of magnitude,
the Newtonian fluctuations of the mutual angle between two separated 
plumb lines (suspended mirrors) produced by atmospheric perturbations; 
preliminary results were presented in \cite{c9}.

Stochastic density variations and transportation of large air masses produces
a contribution into the gravity force on the Earth surface.
Almost all processes of such kind are connected with the corresponding pressure 
changes. Thus they can be controlled by means of pressure monitoring. 
In the gravimetry theory, it is well known that the influence of atmospheric 
pressure on the gravity consists of direct attraction of the atmospheric 
mass and crust deformation due to atmospheric loading \cite{c10}.

The both effects have the same order of magnitude but the loading effect,
as a rule, is at least two-five times less then the direct attraction. 
For this reason, below we consider only the direct attraction effect as one 
of main contributions to the atmospheric Newtonian noises.

As a control reference point in our analysis, we bear in mind an
acceptable level of the noise spectrum density which allows one
to register the inner core oscillations with amplitudes of the
order of 1 m --- the expected signal can achieve $10^{-13}-10^{-14}$ rad 
for the interferometer base equal to 3 km \cite{c2,c5}. Suppose that the period
and quality factor of the inner core of the fundamental mode are 
$\tau_{0} = 3.3\,h$ and $Q=10^{2}-10^{3}$, respectively \cite{c11}. Then one
can take the measurement time $\tau_{m}=10^{6}$ sec which results
in the following angular noise spectrum limit
$\delta \alpha_{f} = 10^{-11}~$rad$\cdot\mbox{Hz}^{-1/2}$.

\section{Plane atmosphere}

A very rough and simple Earth atmospheric model is a half-space filled 
by air stratum of the homogeneous density $\rho$ of the altitude $h$. 
A change in the gravity acceleration $\Delta g$ caused by perturbation of 
the atmospheric density can be estimated as follows \cite{c12}
\begin{equation}
\Delta g_z= \frac{2\pi G}{g}\,\Delta p,
\label{eq1}
\end{equation}
where $\Delta p$ is the atmospheric pressure variation, $g$ is the gravity
 acceleration, $G$ is Newtonian gravity constant and $z$-axis is directed 
 upward. The factor of admittance between the gravity and pressure
$k_{p}=2\pi G/g=0,427\, \mu \mbox{Gal}/\mbox{hPa}$\, is widely used
in geophysics and is in a fair correspondence with experimental data.
One can easily get (\ref{eq1}) starting with 
the surface field of the stratum $g_z=2\pi G {\rho} h$ and using a 
hydrostatic coupling between the pressure and density $p={\rho}gh$ if one
supposes very slow variations.

Although this formula has an integral sense, one can likely use it in a
spectral form by substituting the corresponding spectral densities of the 
gravity and pressure variables, having in mind the fact that the changes of 
pressure are mostly quasistatic processes, i.e. the factor of admittance 
does not depend on frequency.

After this, the relative angular deflection of two plumb lines located 
at the ends of the interferometer's base can be evaluated phenomenologically,
in view of the following expression:
\begin{equation}
<\delta\alpha>\simeq \xi \frac{\Delta g_x}{g}\,\frac{L}{L_{p}}=
\xi\,\frac{k_{p}}{g}\,\frac{L}{L_{p}}\,|p(f)| \sqrt{\Delta f},
\label{eq2}
\end{equation}
\noindent where $L$ is the arm's length, $L_{p}$ is the scale of the
pressure anomaly domain (cyclone size, etc.), 
$p(f)$ is the spectral density of the presure amplitude 
in the bandwidth $\Delta f$ and $\xi \ll 1$ is a
coupling factor between vertical $g_z$ and horizontal $g_x$ components of
gravity variations.

To evaluate the $\xi$ factor, we suppose that pressure changes along the 
$x$-direction so slowly, that (\ref{eq1}) remains to
be valid and it leads to the condition $h\ll L_p$. Then using the potential
character of the gravity field, i.e., ${\rm rot}\ {\bf g} =0$, one can take
$\partial g_x / \partial z = \partial g_z / \partial x$, or
$\Delta g_x / h \simeq \Delta g_z / L_p$. This results in the
estimation
$$
\xi=\frac{\Delta g_x}{\Delta g_z} \simeq \frac{h}{L_p} \ll 1
$$
which might be used while calculating in view of formula (\ref{eq2}).

The pressure time variation background was investigated by many 
meteorological observatories in different places. 
In spite that it is a local characteristic, there are typical common features 
in the spectrum density such as the diurnal $(1.17\cdot 10^{-5}$ Hz) 
and semidiurnal $(2.33\cdot 10^{-5}$ Hz) peaks, a general upward slope at 
very low frequencies, etc.
For numerical estimations at the order of magnitude, we used data collected
by the Mizusawa Astrogeodynamic Observatory (Japan) from 1986 to 1987.

Available data \cite{c13} can be approximatelyly separated into three following
spectral domains with the dominant spectral amplitudes, namely,
$$
\begin{array}{ccc}
(a)& (2\cdot 10^{-5}\,-\,2.5\cdot 10^{-5}),& |p(f)|=
9\cdot 10^{2}\ \mbox{mBar/Hz}^{1/2} \\
(b)& (1,5\cdot 10^{-5}\,-\,8\cdot 10^{-6}),& |p(f)|
=1,9\cdot 10^{3}\ \mbox{mBar/Hz}^{1/2} \\
(c)& (3\cdot 10^{-6}\,-\,5\cdot 10^{-7})  ,& |p(f)|=9,5
\cdot 10^{3}\ \mbox{mBar/Hz}^{1/2}.
\end{array}
$$
This demonstrates that a transition from time-scale of several
hours (a) to time-scale of several days (b) is accompanied by 
the increase in the pressure fluctuations, in average, of one order 
of magnitude. One can see that the domain of several hours (a) 
overlaps with the tidal semidiurnal spectrum peak, which can be removed 
while data processing, so it is reasonable to take for estimation 
a decreased (but still dominant) extrapolation value 
$1\cdot10^{2}\,\mbox{mbr}/\mbox{Hz}^{1/2}$ for the times $3\,\div \,4$ 
hours (inner core polar mode). Then
taking as typical average pressure anomaly the (cyclone) size 
$L_{p}=500\,\mbox{km}$ and the effective atmosphere altitude $h$=10 km,
one obtains the following result for the standard angle deviation 
between arm's mirrors
\begin{equation}
<\delta\alpha>\ \simeq\ \frac{k_p}{g}\ \frac{hL}{L_p^2}\ |p(f)|
\sqrt{\Delta f} \ \simeq \ 5\cdot 10^{-12} \sqrt{\Delta f}\ \mbox{rad}.
\label{eq3}
\end{equation}
\noindent
This estimation does not exceed the critical noise density mentioned in 
Introduction although it shows difficulties in the detection of the 
inner core motion because the angular noise is at the border of the expected 
gravity effect. However, it is supposed that a correction for the
gravity atmospheric noise could be carried out if one would precisely
control pressure variations in the location of front and end mirrors.

\section{Cyclone dynamic effect}

An obvious lack of the previous consideration is an uncertainty of the $\xi$
factor which cannot be regorously found within the framework of the homogeneous 
static model of the atmosphere. To avoid this uncertainty, it would be useful 
to consider a more complex dynamic model where some spatially limited domain 
inside the atmospheric half space, called below as ``cyclone'' (characterized 
by perturbed parameters of density and pressure) moves along the 
interferometer base.
Then a direct calculation of its Newtonian attraction permits one to get a more
reliable estimation of mutual angular deflections of the interferometer arm's
mirrors.

It is convenient to take the cyclone form as an oblate ellipsoid of rotation
because its gravitational field can be expressed in terms of
elementary functions \cite{c14}. After this, the problem is formulated as follows.
The cyclone (ellipsoid of rotation) with its plane of symmetry on the Earth
surface moves along the $x$ axis where the interferometer arm is located
(figure 1). Only upper half of the ellipsoid has a physical sense
representing the cyclone, so that an effective gravitational field along 
the $x$- axis must decrease twice. A vertical size of the cyclone (a minor 
ellipsoid semiaxis $c$) is approximately equal to the atmosphere altitude
in $z$ direction $(c\leq h)$,  however, it is much less then a horizontal 
 size of the cyclone, a major semiaxis $a$ $(a\gg c\sim h)$.  Let us take
a Cartesian coordinate system in the cyclone center in such a way that 
the current value $x$ is just the distance between the cyclone and 
interferometer centers. The positions of the front and end 
suspended mirrors are $(x-L/2)$ and $(x+L/2)$, with
$L$ representing the interferometer arm length. The cyclone pass 
through the interferometer base corresponds to the $x$ variation in
the limits $(+\infty;-\infty)$.

For a homogeneous ellipsoid, its gravity field (acceleration) along the 
$x$ axis grows linearly with the distance from the center inside the 
figure $(x \leq a)$, while outside $(x \geq a)$ it can be described 
by the following expression
\begin{equation}
g_{x}=-2\pi\,G\rho\,\frac{a^{2}c}{A^2}\,\Big(\eta ^{-1}\,\sqrt{\eta^{2}-1}-
\eta\,\arcsin{\eta^{-1}}\Big),
\label{eq4}
\end{equation}
\noindent
where $A^{2}=a^{2}-c^{2}, \,\,\eta=x/A$ and $\rho$ is an excess
(or deficit) of air density over the average density of unperturbed
atmosphere.

In principle, formula (\ref{eq4}), in combination with the corresponding
formula inside the figure, gives a possibility to determine dynamic
perturbations of mirror's plumb lines when the cyclone is penetrating 
through the base. However, it is much more convenient to operate with the field 
of a thin ellipsoid shell to avoid a necessity of the transition between 
two different field expressions.

To get formula for the shell field, it is enough to accept $c=ka$ and then
to differentiate $g_{x}$ (\ref{eq4}) as a function of the variable $a$. This
procedure results in the following formula:
\begin{equation}
dg_{x}=4\pi\,G\rho\,k\,\frac{a^{2}da}{x\sqrt{x^{2}-a^{2}(1-k^{2})}}.
\label{eq5}
\end{equation}
Expression (\ref{eq5}) presents the gravitational field outside the shell
$(x \geq a)$,\, meanwhile, the internal field is equal to zero, the factor
 $da$ is the shell's thickness in the cross point with the $x$ axis.
Then the field of inhomogeneous cloud at the points of $x$ axis can be
calculated in terms of the integral
\begin{equation}
g_x=4\pi\,G\,k\,\int_0^x \frac{a^2\rho(a)\,da}{x\sqrt{x^{2}
-a^{2} (1-k^{2})}},
\label{eq6}
\end{equation}
\noindent
where the function $\rho(a)$ defines the cyclone density distribution (the
maximum horizontal size of the cyclone, hereinafter, noted as $a_0$).

The mutual angular deflection of mirror's suspensions located at the ends of
the interferometer base is given by the simple combination of the function
$g_x(x)$ in two particular points of the $x$ axis:
\begin{equation}
\delta \alpha = \frac{g_x(x+L/2)- g_x(x-L/2)}{2g}.
\label{eq7}
\end{equation}
Factor 2 in the denominator reflects the effect of the upper part of
the ellipsoid cloud only, i.e. the domain where $z \geq 0$. For transition to
the time domain, one has to substitute $x=vt$, where $v$ is the cyclone
velocity.

To illustrate a general type of the angular deflection dynamics, formulas
(\ref{eq6}) and (\ref{eq7}) were calculated numerically for several 
artificially selected laws of the density variation over the cyclone. 
Four types of the density distributions were considered, namely, 
``step,'' ``circle,'' ``harmonic'' and ``Gaussian''
(see figure~2, curves 1a--1d) with the same normalization of mean 
value and variance:
$$\int\rho(x)kx^2dx=(\pi/2)\bar\rho a_0^2c,\qquad
\int\rho(x)kx^4dx=(\pi/2)\bar\rho a_0^4c.$$

Corresponding plots are presented in figure~2.
Density distributions used for calculations are presented by curves 1a--1d
and curves 2a--2d show the dynamics of relative variations of the 
mirror's plumb lines. 
For numerical calculations, it was used $v$=1 m/sec and the other parameters 
were $a_0$=300 km, $c$=10 km, $\bar\rho$=0.02 ${\rm kg/m^3}$ and $L$=3 km.

Figure~2 demonstrates a strong dependence of the mutual angular deflection
on the density distribution inside the cyclone. Common features are peaks of
deflection at the cyclone base borders and a
suppression of perturbation inside the cyclone. The amplitudes of peaks
depend on the density structure of the cyclone cloud. 
The homogeneous cloud produces the biggest jump of deflection (case a).
The Gaussian distribution minimizes the jump effect (case d). 
However, the best compensation of perturbations inside the cyclone takes 
place just for the homogeneous model (here the mutual deflection is more 
close to zero then in the other cases).

For the homogeneous model (case a), the integral in (\ref{eq6}) can be
calculated analytically. Results of such calculation within the first order
accuracy of the parameter $(c/a_{0})\ll 1$ read
$$
g_{x}(x)\simeq - \pi^2\rho\,G\,\frac{c}{a_0}\,x \ \ \ \ 
\ \ \mbox{for}\ \ |x|\leq a_0;
$$ $$
(\partial g/\partial x)\simeq \pi^2\rho\,G\,\frac{c}{a_0}\ \ \ \ 
\ \ \mbox{for}\ \ |x|\leq a_0;
$$ $$
\Delta{g}=g_{x}(x+L/2)\ -\ g_{x}(x-L/2)\simeq g_x|_{x=a_0}
\ \frac{L}{a_0}.
$$
For the mutual angular deflection, one has
\begin{equation}
\delta\alpha \simeq \frac{\Delta g}{2g}\simeq \frac{\pi^2\rho\,G\,c}{2g}\,
\frac{L}{a_0}=
\frac{\pi}{4}k_p\ \frac{c}{v_s^2}\ \frac{L}{a_0}\,\Delta p. 
\label{eq8}
\end{equation}
The adiabatic coupling between the pressure and density variations
$(\rho/\rho_0)=(\Delta p/p_0)$ was used to obtain relationship~(\ref{eq8}), 
where $v_s^2\simeq p_0/\rho_0$ is the sound velocity in the normal atmosphere
and $\Delta p$ is the pressure jump outside the cyclone.
By comparing (\ref{eq8}) with phenomenological formula (\ref{eq3}), one
can see that the formulas are similar.
For $c\simeq h\simeq$10 km, $(h/v_s^2)\simeq 1/g$
($v_s$=330 m/sec) one obtains $\xi\sim 1$ for the rough
homogeneous model, but it decreases up to 0.01 for more smooth cyclone
shapes.

From figure~2 it is clear that a more favorable
cyclone structure can be presented by a cloud with Gaussian fronts and
long plane central zone. Then jump perturbations corresponding to the 
cyclone front passing through the interferometer base are suppressed as
well as perturbations in the central part of the cloud. In general,
large sharp perturbations (corresponding to relatively rare events) 
can be obviously controlled by barometric measurements. At quiet atmospheric
conditions, the usual daily variation is of the order of $(1 \div 10)$ mBr, 
that keeps the effect of Newtonian
atmospheric attraction, in view of (\ref{eq8}), at the level
of $10^{-11}\div 10^{-12}$ rad.

Nevertheless, it is by two--three order of magnitude larger that the inner
core oscillation effect $(\sim 10^{-13}\div 10^{-14}$ rad). However,
(\ref{eq8}) gives the integral value of the atmospheric gravity effect
and one has to evaluate its spectrum noise density in the domain of interest
at $f\simeq (10^{-4}\div 10^{-3})$ Hz.

Let us consider Poissonian flux of cyclones with average number $\bar n$
per time unit. Excluding the cases of recovering, one can take
$\bar n \le v/2a$. The well-known formula for the power spectrum of
the Poissonian flux defined by the spectrum of individual events reads
\begin{equation}
S_\alpha(f)=\bar n|\delta\alpha(f)|^2,
\label{eq9}
\end{equation}
where $\delta\alpha(f)$ is the Fourier component of the angle deflection.
It is clear {\it a priori} that the spectrum maximum takes place near
$f \sim v/a=10^{-6}$ Hz independently of the cyclone density
distribution. The spectrum behavior at high frequencies strongly depends
on the cyclone density structure. The main role is played by the peak amplitude
at the border $x=vt=a$. For the thin ellipsoid shell, the peak amplitude 
is $4\pi G\rho da$ and it gives the elementary gravity variation
spectrum
$$
dg_x(f)=\frac{4\rho Gv}{f} \cos\frac{2\pi af}{v}\,.
$$
It has to be integrated over the cyclone cloud in order to get the 
total gravity spectrum
\begin{equation}
g_x(f)=\int_0^{a_0}dg_x(f)=
\frac{4Gv}{f} \rho\left(\frac{2\pi f}{v}\right),
\label{eq10}
\end{equation}
i.e. it is essentially determined by the cloud density spectrum $\rho$. 
To obtain the mutual angular spectrum of plumb lines, one has to multiply
(\ref{eq10}) by the transfer function of the differential link (\ref{eq7}).
Numerical calculations for the four selected cyclone clouds are presented
in figure~2 by curves 3a--3d.

Let us consider the homogeneous cyclone (case a) in more detail. 
The integrand in (\ref{eq10}) is calculated analytically and the final 
result for the angular deflection power spectrum reads
\begin{equation}
\delta\alpha(f)\ =\ \frac{4\rho G v^2 \sqrt{\bar n}}{\pi f^2}
\ \sin\left(\frac{2\pi a}{v}f\right)
\ \sin\left(\frac{\pi L}{v}f\right)
\ \sqrt{\Delta f}.
\label{eq11}
\end{equation}
At $f>v/L\simeq 10^{-4}$ Hz, the spectrum oscillates and falls
down as $1/f^4$ and at low frequencies $S_\alpha\sim 1/f^2$.
At frequency $f\sim 10^{-4}$ Hz (interesting for us), 
the angular spectrum density equals to 
$2\cdot 10^{-9}~$rad$\cdot\mbox{Hz}^{-1/2}$, 
that is two order of magnitude larger than the acceptable level. 
However, it follows from the very sharp
density jump in the homogenous cyclone model. For more smooth cyclone
structure, the angular spectrum at high frequencies goes down much faster:

$S_\alpha\sim 1/f^3$ (3b);

$S_\alpha\sim 1/f^4$ (3c); 

$S_\alpha\sim\exp{-f^2}$ (3d). 

From figure~2 one can see, that at frequency $10^{-4}$ Hz the amplitude 
of angular noise is already less than the critical level 
$10^{-11}~$rad$\cdot\mbox{Hz}^{-1/2}$ (3c) and 
then
it becomes negligible (3d).

Analysis presented in sections 1 and 2 elucidates particular characteristics
of the free mass interferometer such as a gravity gradiometer one. Being a
gravity device, it is affected by all environmental movements of
masses. But its transfer function suppresses the influence of large scale
homogeneous gravity sources by the factor of the ratio of the base length 
to the source scale.
The most effective sources would be those which generate the gravity field
with a space scale equal to the interferometer base (or less).
Such sources might have a typical variation time of the order of tens seconds
for $v\ge 10$ m/sec, so it would be high frequency sources with respect to
the main geophysical processes with times equal to hours and longer. 
It is clear also that, by suppressing the Newtonian environmental
noise, the gravity gradiometer suppresses equally low frequency (quasistatic)
signals arising from the objects
(including the inner core oscillations). However, it cannot be a serious 
objection if instrumental (intrinsic) noises of the device are enough small 
to permit measurements of such very weak effects.
It seems that modern advanced technology of gravitational interferometers
under construction could meet these requirements.

\section{Turbulent flux efect}
 
The previous consideration was carried out at the condition 
of the hydrostatic atmosphere or, at least, slow laminar air currents.
However, it is known that in upper troposphere strata the Reynolds
number can be very high Re$ > 10^{6}$ and appearance of turbulent currents
is a typical event. The rigorous calculation of the turbulent atmosphere
gravity effect is complicated due to an uncertainty in constructing the
models adequate to our problem, many unknown parameters
and general complexity of the theory of turbulent atmosphere.
For the estimation of the order of magnitude of the gravity noise produced 
by the air turbulence, we hope that a simple phenomenological consideration 
based on the Kolmogorov--Obukhov (K-O) law \cite{c15} can be useful.
This law fixes relative variation of velocities
of vortex zones with different scales in the regime of developed turbulence.

Let $a$ be some maximum size of the turbulence zone with an average
amplitude of the velocity variations $V$. Then according to the K-O law,
 a velocity
standard  $v_{\lambda}$ at the small scale $\lambda \leq a$ obeys to the
formula
\begin {equation}
 v_{\lambda} \simeq V\left(\frac{\lambda}{a}\right)^{1/3}.
\label{eq19}
\end{equation}
For each $\lambda $, 
one can define the frequency of the velocity fluctuation as 
$f=V/\lambda$. Then (\ref{eq19}) can be rewritten as 
\begin{equation}
v_{\lambda}= V^{4/3}(f\,a)^{-1/3}.
\label{eq20}
\end{equation}
Following \cite{c16} the pressure standard $\Delta p$ is introduced in
(\ref{eq10}) through the hydrodynamic law at the scale $\lambda$:
$$\Delta p_\lambda\simeq\frac{\rho\, v_{\lambda}^{2}}{2}\,,$$ 
while the variance $<(\Delta p)^{2}>$ is composed by contributions 
of all turbulent scales $\lambda$, or corresponding frequencies 
$f(\lambda)$ from a minimal scale,
say, $\lambda_{0}$ up to the current value $\lambda$, i.e.
\begin{equation}
 <(\Delta p)^{2}>=\int_{f(\lambda)}^{f(\lambda_{0})}\,S_p(f)\,df\,\,\simeq
\mbox{const} - \int_{0}^{f(\lambda)}\, S_p(f)\,df.
\label{eq21}
\end{equation}
Here $S_p(f)$ is the spectral power density of the pressure variations
and the approximation $ \lambda_0\simeq 0$, $f(\lambda_0)\simeq\infty$ was
used.

Differentiating (\ref{eq21}) over frequency and taking into account
(\ref{eq20}), one obtains the well-known spectrum of developed turbulence
\begin{equation}
S_p(f)\simeq \left(\frac{\rho}{3}\right)^2\,V^{16/3}\,a^{-4/3}\,f^{-7/3}.
\label{eq22}
\end{equation}
The frequency domain where the spectrum (\ref{eq22}) is valid
reads
$$ 
\frac{V}{a}=f_{2}\, \leq \,f \,\leq  f_{3}=\frac{V}{a}\,\mbox{Re}^{3/4}.
$$ 
At frequencies $ f<f_{2}$ (the domain of slow hydrodynamic changes), the
pressure variations conventionally obey to the $\Delta P \sim f^{-1}$ law 
(velocity fluctuations $v_{\lambda}$ are considered as independent), so the
pressure spectrum in this domain reads
\begin{equation}
S_p(f)\simeq \, \frac{d\,(\Delta P)^{2}}{df}=\,S_0\,f^{-3}.
\label{eq23}
\end{equation}
Here the constant $S_0$ can be determined by the condition of uninterrupted
spectra in the point $f_{2}$ for both approaches under consideration. 
It is obvious also that the lowest conceivable frequency associated 
with the limited turbulence zone is $f_{1}=V/R_{e}$, where $R_{e}$ is the 
Earth radius. Below this point, the spectrum of pressure variations 
remains to be uncertain at least within the framework of the approach used.

This simple model of the turbulent atmosphere pressure spectrum
does not take into account all complex processes which could take place.
So, experimental observations of thunder storms in the frequency domain
 $(10^{-2}\div 10^{-4})\,$ Hz gives the spectrum indices distributed, in 
 average, between 2 and 4.25 and centered in the point 3.5 \cite{c17}, i.e.
 larger than the K-O index $7/3$. 
On the contrary, at frequencies below $10^{-4}$, experimental data 
better agree with index $5/3$ \cite{c18}, i.e. low frequency fluctuations
grow slowly than it is predicted by the flicker type noise with 
the spectrum index 3, and so on.

Nevertheless, this model at the quiet atmosphere conditions corresponds
to the real experimental data by the order of magnitude.
In particular, it was successfully 
used for calculating the atmospheric load deformations measured by strain 
meters at the Obninsk Seismic Station (see \cite{c16}). In figure~3 (adopted from
\cite{c16}), one can compare the model spectrum calculated for the turbulent
zone with kinetic parameters $ a=L=18$ km, $V=\Delta U=3$ m/sec and average
density $\rho_0=0.55\ \mbox{kg}/\mbox{m}^3$ (corresponding altitude $h=8$
km) with the experimental pressure spectrum measured at the Obninsk Seismic
Station. One can see a satisfactory coincidence, at least, by the order of
magnitude. 

Now we discuss the gravity noise associated with a turbulent atmospheric
process. For this, let us consider the plane atmosphere with effective 
altitude $h$ filled by the turbulent flux as it was considered
in section 2. The estimation of gravity variations on the Earth surface 
is taken as follows $$\Delta g\simeq 2\pi Gh\Delta\rho.$$ 
The average density changes $\Delta\rho$
over the total scale of turbulent zone can be replaced by the average
pressure standard deviation using the adiabatic law, i.e.
\begin{equation}
\Delta g\simeq 2\pi G\frac{\rho_0 h}{p}\Delta p.
\label{eq24}
\end{equation}
The mutual plumb lines deflection has to be a function of the pressure
gradient, i.e.
\begin{equation}
\delta\alpha\ \simeq\ \frac{1}{2g}\ \frac{\partial\Delta g}{\partial x} L
\ =\ \frac{1}{2g} 2\pi G\frac{\rho_0 h}{p^2}
\left(\frac{\partial p}{\partial x}L\right) \Delta p.
\label{eq25}
\end{equation}
(In this approximation, we suppose that the standard pressure deviations do
not depend on $x$).
To evaluate the pressure gradient along the interferometer base, one can
employ again the K-O law or use direct experimental data provided by some
spatial barometric grid. Below we consider both methods.

One can express the pressure gradient along a turbulence cell 
($\lambda\sim L$) in terms of the spatial derivative of the local relation
$p = \rho_0 (v_{\lambda}^{2}/2)$
\begin{equation}
\frac{\partial p}{\partial x}=\,\rho_0 v_{\lambda}\,
\frac{\partial v_\lambda}{\partial x}\ = L\rho_0\,
\frac{v_\lambda}{L}\frac{\partial v_\lambda}{\partial x}.
\label{eq26}
\end{equation}
If $v_\lambda$ is not known, one can take it from the K-O law, $v_\lambda=
V(L/a)^{1/3}$, and then, in view of the linear dependence of the
velocity on $x$,
i.e. $(v_\lambda/L)\simeq (\partial {v_\lambda}/\partial x)$, one obtains 
\begin{equation}
\frac{\partial p}{\partial x}\,\simeq L\,\rho_0\left(\frac{V}{L}\right)^{2}
\left(\frac{L}{a}\right)^{2/3}\,\,\,\sim 10^{-2}\ \mbox{hPa/km},
\label{eq27}
\end{equation}
where the parameters in figure~3 used for numerical estimation are
$a$=18 km, $V$=3 m/sec and the interferometer base $L$=3 km.

This very approximate estimation is supported by experimental data provided by
the barometric space pattern covered several hundred kilometers around the
Brussels \cite{c10}. Measured value of the pressure gradient along the
East--West direction was in average 0.037 hPa/km (and 0.085 hPa/km in 
the Nord--South direction). This (by the order of magnitude) is in a
satisfactory agreement with theoretical calculation. Below, we use the
intermediate value $5\cdot 10^{-2}$ hPa/km for estimations.

To transform formula (\ref{eq25}) for the relative angular variations
of arm mirrors, one can substitute $(\rho_0/p_0)=v_s^2$ and use the
K-O spectrum (\ref{eq22}) for estimating the standard deviation $\Delta p$.
It results in the formula for angle deflection
\begin{equation}
<\delta\alpha(f)>\ \simeq\ \frac{\pi Gh}{pgv_s^2}
\frac{\partial p}{\partial x}L\ <S_p(f)>^{1/2} \sqrt{\Delta f}.
\label{eq28}
\end{equation}
For numerical estimations, let us take the following parameters: 
$v_s$=330 m/sec, the average pressure magnitude $p=10^3$ hPa, 
the pressure jump at the arm ends 
$$\Delta p_L=\frac{\partial p}{\partial x}\, L=0.05~\mbox{hPa/km}\cdot 
 3~\mbox{km} =0.15~\mbox{hPa},$$
  and the pressure spectrum density $$S_p(f=0.001\mbox{Hz})\simeq 10^3
\ \mbox{Pa}^2/\mbox{Hz}.$$ As a result, one obtains $<\alpha(f)>\simeq
2\cdot 10^{-14}~$rad$\cdot\mbox{Hz}^{-1/2}$.
The noise is proved to be small enough to be neglected in measurements of
the Earth core gravity oscillations.

It is worthy to estimate a residual Newtonian angular
noise in the frequency domain typical for gravitational wave detectors, i.e.
above 10 Hz. Following the K-O law the numerical result must be multiplied
by the factor $(10\ \mbox{Hz}/0.001\ \mbox{Hz})^{-7/5}\ =\ 2.5\cdot 10^{-6}$,
that yields 
$<\alpha(10\ \mbox{Hz})>\ \simeq\ 5\cdot 10^{-20}~$rad$\cdot\mbox{Hz}^{-1/2}$
(the frequency 10 Hz just corresponds to the upper boundary of validity of
the K-O approach if Re$\geq10^6$, $V$=3 m/sec, $a$=20 km).

For VIRGO superattenuator, a length of suspension is of the order of 10 m, it
provides the interferometer optical arm variations on the level of $5\cdot
10^{-17}\ \mbox{cm}/\mbox{Hz}^{1/2}$. This means that there exists a 
deformation noise at $1.7\cdot 10^{-22}\ \mbox{Hz}^{-1/2}$, which might
create some problems for advanced VIRGO project.

\section{Simulation of plumb lines deflection}

To test a likelihood of our estimations, we carried out a simple
computer simulation of the angular dynamics of two separated plumb
lines (pendula) in a variable pressure field. Real experimental
data were kindly provided by Japanese Weather
Association. These data were collected by the Meteorological National
Geographical Institute at the area close to Tsukubo Scientific Center.
The area 20x20 km was covered by 10 meteostations with an average
intermediate distance 10 km. The pressure data were
registered at each station with a sample time 1 min.
An accuracy of measurement was 0.01 $\%$.
The recorded signal contained obvious diurnal oscillations of the pressure;
the average value of these oscillations was 0.1 $\%$ of atmospheric
pressure ($\sim$100 Pa), meanwhile, the differential pressure amplitude 
for neighbour stations had only 10 $\%$ accuracy.

For modeling the angular free-mass interferometer response, we used
the simple hydrostatic approach calculating the mutual angle between plumb
lines of suspensions according to the formula
\begin{equation}
\Delta\alpha \sim 2\pi\frac{GhL}{gv_s^2}\,\frac{p_i-p_k}{|r_{ik}|},
\label{eq30}
\end{equation}
where $p_i$ is the pressure at the $i$-station, $r_{ik}$ is the 
distance between neighbour stations (a linear interpolation 
$p(\kappa x)=\kappa p(x)$, $\kappa\le 1$ was used).  In figure~4 (a,b), 
the spectrum of mutual angular fluctuations for pendula separated by the
 distance $L$ is shown. It demonstrates the flicker noise at zero
 frequencies, diurnal peak and a moderate noise level 
 $\leq 10^{-11}~$rad$\cdot\mbox{Hz}^{-1/2}$ at times over few hours. 
 In fact, real noise can be less because the noise level in figure~4 
 is determined just by the unsufficient accuracy of the pressure 
 measurements.

\section{Conclusions}

Our estimations as well as computer simulation of the pendula
behaviour in real variable pressure field show that, under normal
atmospheric conditions, the mutual angle between two suspensions spatially
separated by 3--4 km fluctuates with the standard spectral deviation of the
order of $10^{-11}~$rad$\cdot\mbox{Hz}^{-1/2}$ in the frequency domain 
$10^{-3}\div 10^{-4}$ Hz, although some rare events (cyclones, storms, etc.) 
can produce one--two order larger perturbations. Such events, however, 
can be forecasted in advance, registered and removed from standard data 
records, or simply the measurements must be eliminated in such times.
Thus it is quite possible to achieve the desirable
accuracy $\sim 10^{-13}\div 10^{-14}$ rad for the measurement time of
 $10^4\div 10^6$ sec for detecting the inner core motion.
 
Our rough approximations of gravity noises for the turbulent atmosphere
have shown that the high frequency tail of Newtonian atmospheric
fluctuations can provide some difficulty for gravitational wave detection
at very low frequencies $f\leq 10$ Hz. This point, however, requires a more
detailed investigation with more adequate models of turbulent currents in the
high atmosphere.

\section*{Acknowledgments}

Authors appreciate kind assistance of the Japan Meteorological Agency,
Mr. Yuzo Arisawa and Dr. Akihiko Itagaki for providing observational data
collected by the Tsukubo Meteorological Station.

Authors express their gratitude to Prof. Isao Naito and Prof.
Masa-Katsu Fujimoto from NAO, Japan for their interest to this work
and discussions.

We got many useful information and critical remarks from our
collegues: Prof. G. S. Golitzin, Prof. S. N. Kulichkov from the
Atmosphere Physics Institute of the Russian  Academy of Sciences 
and Prof. A. B. Manukin and Prof. Yu. A. Gusev from the Earth
Physics Institute of the Russian Academy of Sciences.

This work was partially supported by RS Grant (UK) No RCPX331.

\section*{Figure captions}

Figure~1. Effect of the cyclone dynamics.

Figure~2. Density distribution over the cyclone (1),
``step'' $\rho=$const for $x\leq x_0$ (1a), 
``circle'' $\rho \sim \sqrt{1-x^2/x_0^2}$ for $x\leq x_0$ (1b), 
``harmonics'' $\rho \sim \cos(\pi x/x_0)$ for $x\leq x_0$ and 
``Gaussian'' $\rho \sim \exp(-x^2/x_0^2)$ for all $x$,
 angular response as function of time (2) and its power spectum (3). 
 
 Figure~3. The pressure spectrum on the Earth surface.
 Calculated curve with $h=18~$km, $V=3~$m/sec and $\rho=0.55~$kg/m$^3$ (a)
 and experimental data (b). 

Figure~4. Spectral density of angular response calculated using data of 
the Japan Geographic Institute (Tsukubo Metereological Station).
Nord--South direction (a) and East--West direction (b).

\end{document}